\documentclass[twocolumn,aps,prb,english,twocolumn,epsfig,groupedaddress]{revtex4}
\usepackage{epsfig}
\usepackage{amsmath}
\usepackage[latin1]{inputenc}
\usepackage{latexsym}
\usepackage{ulem}
\usepackage{color}

\begin{document}
\title{On the reliability of recent Monte Carlo studies of dilute systems of localized spins interacting with itinerant carriers.}
\author{Richard Bouzerar$^{1,3}$ and Georges Bouzerar$^{2,3}$}
\affiliation{
$1.$ LPSC, Universit\'e de Picardie Jules Verne, 33 rue Saint-Leu, 80039 Amiens Cedex 01 \\
$2.$ Institut N\'eel, CNRS, d\'epartement MCBT, 25 avenue des Martyrs, B.P. 166, 38042 Grenoble Cedex 09, France \\
$3.$ Institut Laue Langevin, Theory Group, B.P. 156 38042 Grenoble Cedex 09, France \\
}
\date{\today}
\begin{abstract}
In this paper, we discuss magnetic properties of dilute systems of localized spins interacting with itinerant carriers.
More precisely, we compare recent available Monte Carlo results with our two step approach (TSA) calculations. The TSA consists first on the determination of the magnetic couplings and then on a proper treatment of the resulting effective dilute Heisenberg Hamiltonian. We show important disagreement between the Monte Carlo results (in principle exact) and our TSA calculations. We analyze the origins and shed light on the reasons of those dissensions.
In contrast to one could expect, we demonstrate that the available MC calculations suffer from severe numerical shortcomings. More precisely, (i) the finite size effects appear to be huge in dilute systems, (ii) the statistical sampling (disorder configurations) was far too small, and (iii) the determination of the Curie temperature was too rough. In addition, we provide new arguments to explain a recent disagreement between the Monte Carlo simulations and TSA for the model study of the well known III-V diluted magnetic semiconductor Ga$_{1-x}$Mn$_{x}$As. We hope that this work will motivate new systematic large scale Monte Carlo calculations.
\end{abstract}
\maketitle

\section{Introduction}

During the last decade, the study of the effects of disorder on magnetic properties (Curie/N\'eel temperatures, magnetic excitations spectrum, spin susceptibility,...) of disordered/dilute magnetic materials became a central field of research for both theoreticians and experimentalists. It is now clear that even for a qualitative understanding of the underlying physics, disorder, dilution and thermal fluctuations should be handled in a proper way. Among widely studied materials one finds oxides manganite as for example La$_{1-x}$Sr$_{x}$MnO$_3$ or La$_{1-x}$Ba$_{x}$MnO$_3$, which exhibit the fascinating giant magneto-resistance phenomenon \cite{Jin,Schiffer,revue-manga1,revue-manga2}. The diluted magnetic semiconductors (DMS) as the well known III-V compound Ga$_{1-x}$Mn$_x$As \cite{VanEsch97,matsukura,Kirby04,Glas04} or the II-VI Zn$_{1-x}$Cr$_x$Te \cite{saito} have attracted a considerable attention. The interest for DMS is partly motivated by their technological potential for spintronics. One can also mention the so-called $d^{0}$ materials as the wide band gap oxides HfO$_{2}$, ZrO$_2$, CaO, ZnO, or the irradiated graphite layers,...\cite{d01,d02,d03,d04,d05}. These new materials may become an alternative for spintronic applications. The Heusler alloys as Ni$_{2+x}$Mn$_{1-x}$Sn \cite{kuebler,Groot} or the double perovskites as Sr$_2$FeMoO$_6$ \cite{dperov,Sarma} are also materials in which disorder clearly plays a crucial role.

Ab-initio based studies have provided the most efficient and reliable tool, to allow theoretical and quantitative study of the magnetic properties without adjusting parameters. As an example, it was possible to study in great details and quantitatively the magnetic properties of the III-V diluted magnetic semiconductor Ga$_{1-x}$Mn$_x$As, both in the presence or absence of additional compensating defects \cite{bergqvist,gbouzerar1,gbouzerar2}. In order to test the interest of a particular material for technological applications, the ab initio approach provides the most appropriate tool. However, because it is material specific and very complex, it does not allow to draw general conclusions which could be valid for a whole family of materials. The study of relevant minimal model is crucial and allows to fill this gap. The model approach is essential to understand the influence of a particular physical free parameter. It can also help for the determination of potential spintronic candidates. The minimal one band model that describes itinerant carriers (holes/electrons) interacting with localized moments reads,

\begin{eqnarray}
H=-\sum_{ij} t_{ij} c^{\dagger}_{i\sigma}c_{j\sigma} + \sum_{i} p_{i}J_{i} {\bf S}_{i}\cdot {\bf s}_{i} +\sum_{i\sigma} \epsilon_{i}c^{\dagger}_{i\sigma}c_{i\sigma}
\label{Hamiltonian}
\end{eqnarray}

In the first term, t$_{ij}$$=$t if i and j are nearest neighbor, otherwise it is 0. The random variable $p_i$ is one if the site is occupied by a magnetic impurity, otherwise it is 0, we denote $x$ the concentration of impurities. ${\bf S}$$_{i}$ denotes the localized impurity spin at site i. For most of the materials S is large enough, we consider here the case of classical spins only. Moreover in ferromagnets the quantum fluctuations are not relevant. In the case of manganites $p_i=1$ (one Mn per site) J$_{i}$ is the Hund coupling (J$_H$) between the itinerant e$_g$ carriers and the localized $t_{2g}$ magnetic moments ($S=3/2$). In diluted magnetic semiconductors, J$_{i}$ is the local p-d coupling (J$_{pd}$). The additional last term describes the effects of the disorder induced by the substitution of a cation by another one. Note that this term includes in particular the electrostatic contribution which originates from the difference of charge between the substituted cation and the original one.

In this manuscript, all the calculations are performed on a simple cubic lattice. We also neglect in eq.(\ref{Hamiltonian}) the last term. The case of finite on-site potential will be discussed elsewhere \cite{ric_inprep}. We expect the more realistic case of a multi-band Hamiltonian to lead to similar conclusions to those that will be presented in this manuscript.

\section{The two steps approach predictions}

Reliable calculations requires several crucial conditions to be fulfilled.
In other words, one should treat (i) the local coupling in a non perturbative way, (ii) the thermal and transverse spin fluctuations properly, and (iii) the disorder and dilution effects should be treated exactly (beyond an effective medium approach). The last condition allows for the localization of the itinerant carriers. This will strongly affect both magnetic and transport properties. The {\it full} Monte Carlo method is {\it a priori} the best procedure, since localized spins degrees of freedom and itinerant carriers are treated exactly, simultaneously and on equal footings. Note that the word {\bf full} is added to avoid confusion with the Monte Carlo treatment of the effective Heisenberg Hamiltonian in the two step approach procedure presented below. An alternative approach which fulfill these requirements is the Two Step Approach (TSA).
TSA is in spirit similar to what is widely used in the ab-initio based studies. It consists in two different stages.
First, for a given configuration of disorder, at $T=0~K$, one diagonalizes the Hamiltonian given by Eq. (\ref{Hamiltonian}).This provides the spin resolved one particle Green's Function of the itinerant carrier (GF) $G^{\sigma}_{ij}(\omega)$ (where $\sigma=\uparrow$ or $\downarrow$). In this first step the appropriate magnetic texture for the localized spins is used. For the ferromagnetism, all localized spins are parallel. Note that, the calculations are done without any approximations at $T=0~K$, thus vertex corrections and the multiple scattering of the carriers on the magnetic impurities are included exactly. From $G^{\sigma}_{ij}(\omega)$ one then calculate the magnetic couplings between the localized moments (integrate out the carriers degrees of freedom). We end up with the disordered/dilute Heisenberg Hamiltonian which reads,
\begin{eqnarray}
H_{Heis}=-\sum_{i,j}p_{i}p_{j}J_{ij}{\bf S}_{i}\cdot {\bf S}_{j}
\label{Heisenberg}
\end{eqnarray}
where the magnetic couplings are given by the well known expression \cite{Lichtenstein84,Katsnelson00}
\begin{eqnarray}
J_{ij}=-\frac{1}{4\pi} \Im m(\int_{-\infty}^{+\infty} f(\omega)\Delta_{i} G^{\uparrow}_{ij}(\omega)\Delta_{j}G^{\downarrow}_{ji}(\omega)d\omega)
\label{couplings}
\end{eqnarray}
In the present case $\Delta_{i} $=$ \Delta_{j} $=$ J_{pd}$ (exchange splitting).
The second step of the TSA consists in diagonalizing reliably the effective disordered Heisenberg Hamiltonian (Eq. (\ref{Heisenberg})). Note that the diagonalization can be performed either by a Monte Carlo treatment (as mentioned above) or within the semi-analytical method called Self-Consistent Local RPA (SC-LRPA)\cite{gbouzerar1}. Once again, to avoid any confusion, we underline and distinguish between the {\bf full} MC study (no mapping to Heisenberg Hamiltonian) and the Monte Carlo that can be used to diagonalize Hamiltonian (\ref{Heisenberg}) in the TSA.
It is important to recall that TSA have already been successfully used for the study of various realistic materials. In these studies, the first step of the TSA  was provided by \textit{ab initio} calculations (realistic exchange integrals). The agreement between calculated Curie temperatures and the experimental data was excellent \cite{Edmonds02,Edmonds04,Bouzerar2,G.Bouzerar06b}. Concerning the second step, it has also been shown that for a given set of exchange integrals, the SC-LRPA is as accurate as Monte Carlo calculations \cite{Bergqvist,G.Bouzerar07,G.Bouzerar07b}. As it will be seen in the following, one great advantage of the TSA is the very large system sizes that can be handled, but are still inaccessible within the full Monte Carlo method. Thus, in this manuscript the second step of the TSA is performed within SC-LRPA.

\begin{figure}[htbp]
\includegraphics[width=9.0cm,angle=-0]{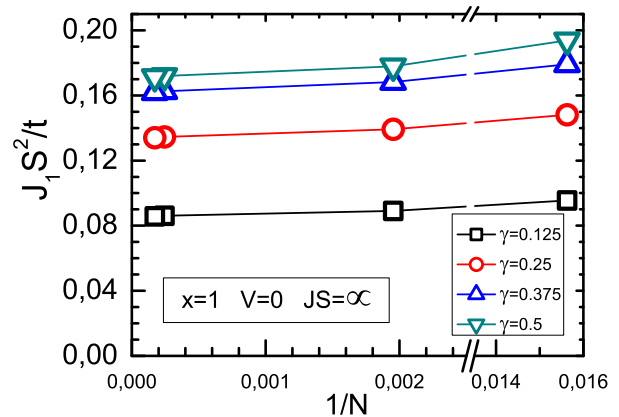}
\vspace{0.0cm}
\caption{(Color online.) Nearest neighbor coupling as a function of 1$/$N (where N is the number of sites) for different $\gamma$ (the hole density is $p_{h}=\gamma x$) in the case of a non dilute (e.g. $x=1$) and non disordered system ($V=0$), in the double exchange limit (JS$=$$\infty$).}
\label{Fig1}
\end{figure}

\subsection{Results for non dilute systems}

Before discussing the case of dilute magnetic systems, it is important to discuss the non dilute case, e.g $x=1$.
A comparison between the full Monte Carlo treatment and the two steps approach has been done in the limit JS$=\infty$ (double exchange (DE) limit)\cite{G.Bouzerar07}. In this case, the nearest neighbor coupling between localized moments reads $J_{ij}S^{2}=t_{ij} \langle c^{\dagger}_{i}c_{j} \rangle$, the spin index $\sigma$ is irrelevant in this case, the fermion operators correspond to the majority spin.
The comparison was performed in the presence of Anderson disorder, e.g the variables $\epsilon_{i}$ in Eq.(\ref{Hamiltonian}) were chosen randomly in the interval $[-$V$/2,$V$/2]$, V denotes the strength of the disorder. Even in the presence of disorder an excellent agreement between the full Monte Carlo calculations and the two step approach was achieved, the difference in the Curie temperature was within 10 $\%$.
\begin{figure}[htbp]
\includegraphics[width=9.0cm,angle=-0]{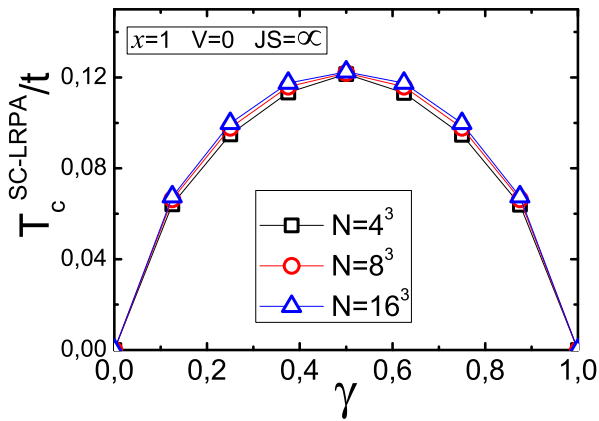}
\vspace{0.0cm}
\caption{(Color online.) Curie temperature (SC-LRPA) obtained within the two step approach as a function of $\gamma$ for different system size, in the case of a non dilute system ($x=1$) and in the double exchange regime (JS$=$$\infty$) and $V=0$. }
\label{Fig2}
\end{figure}
We now show that finite size effects are very small for the non dilute case. This will explain the agreement found between TSA calculations performed on large systems and full Monte Carlo simulations done with much smaller clusters. In Fig.~\ref{Fig1}, the nearest neighbor magnetic coupling (J$_{1}$) between impurities as a function of $1/N$ (N is the number of sites) is plotted. It is clearly observed that J$_{1}$ depends very weakly on the system size. The finite size effects are negligible at any carrier density ($p_{h}=\gamma x$).
In Fig.~\ref{Fig2}, we have plotted the Curie temperature as a function of the carrier density per magnetic moment ($\gamma$).
We recall that in the clean limit the local SC-RPA reduces to the standard RPA calculation. In the case of classical spins the Curie temperature is given by
$T_{C}= \frac{2}{3} (\frac{1}{N}\sum_{\textbf{q}\ne \textbf{0}} 1/E(\textbf{q}))^{-1}$, the magnon excitation dispersion reads $E(\textbf{q})=J_{1}S^{2}(1-\frac{1}{3}(cos(q_{x}a)+cos(q_{y}a)+cos(q_{z}a)))$. We clearly observe that the finite size effects are completely negligible even for very small systems. For the smallest system ($N=4^3$), the Curie temperature is already accurate within less than 2\%. Thus, in the case of clean systems and in the double exchange regime (JS$=$$\infty$), the finite size effects are indeed completely negligible. Similar conclusions have been reached in the case where the on-site disorder ($V\ne 0$ in the previous Hamiltonian)is also included \cite{ric_inprep}.We have even obtained that the average over only few configurations of disorder is enough to provide an accurate value. Thus, one can understand the success of the full Monte Carlo simulations to provide accurate Curie temperature for the study of the non dilute DE model even though in these studies the systems sizes are typically of the order of $N \approx 5^{3}$ sites. The restriction to such small clusters is in fact due to the very large CPU time and memory costs of the standard full Monte Carlo simulations.

\begin{figure}[htbp]
\includegraphics[width=9.0cm,angle=-0]{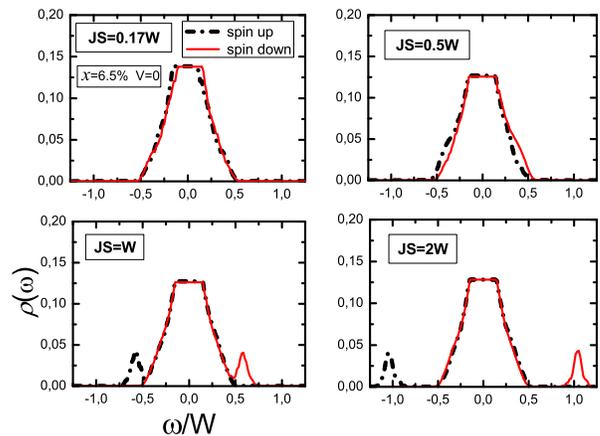}
\vspace{0.0cm}
\caption{(Color online.) Spin resolved Density of states for a fixed concentration of magnetic impurities $x=0.065$ and different values of the local coupling JS. The spin down (resp. up) band is represented by the continuous (resp. dotted) line. Energies are in unit of the bandwidth $W=12t$. }
\label{Fig3}
\end{figure}

However, as it will be seen in the following subsection, for dilute systems, the use of too small system sizes as well as too weak statistical sampling affect the results in a dramatic way.

\subsection{Results for dilute systems}

For simplicity and to allow a direct comparison with the available full Monte Carlo data, we set the impurity concentration to $x=0.065$. We will vary both, the amplitude of J (its sign is irrelevant for classical spins) and the carrier density. Fig.~\ref{Fig3} shows the DOS for various coupling JS. We observe that the impurity band (IB) splits from the valence band for values of $JS \ge W$. In the strong coupling regime, $JS=2W$, the IB is completely separated from the valence band, the magnetic couplings are expected to be dominated by the double exchange mechanism.
\begin{figure}[htbp]
\includegraphics[width=9.0cm,angle=-0]{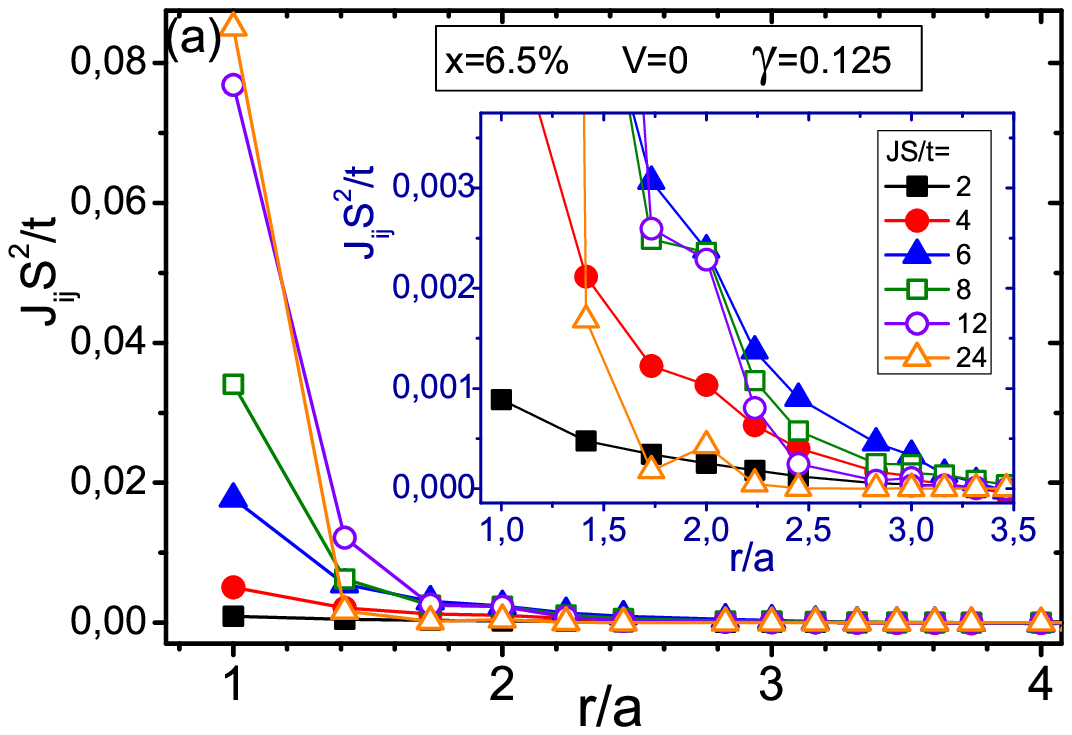}
\includegraphics[width=9.0cm,angle=-0]{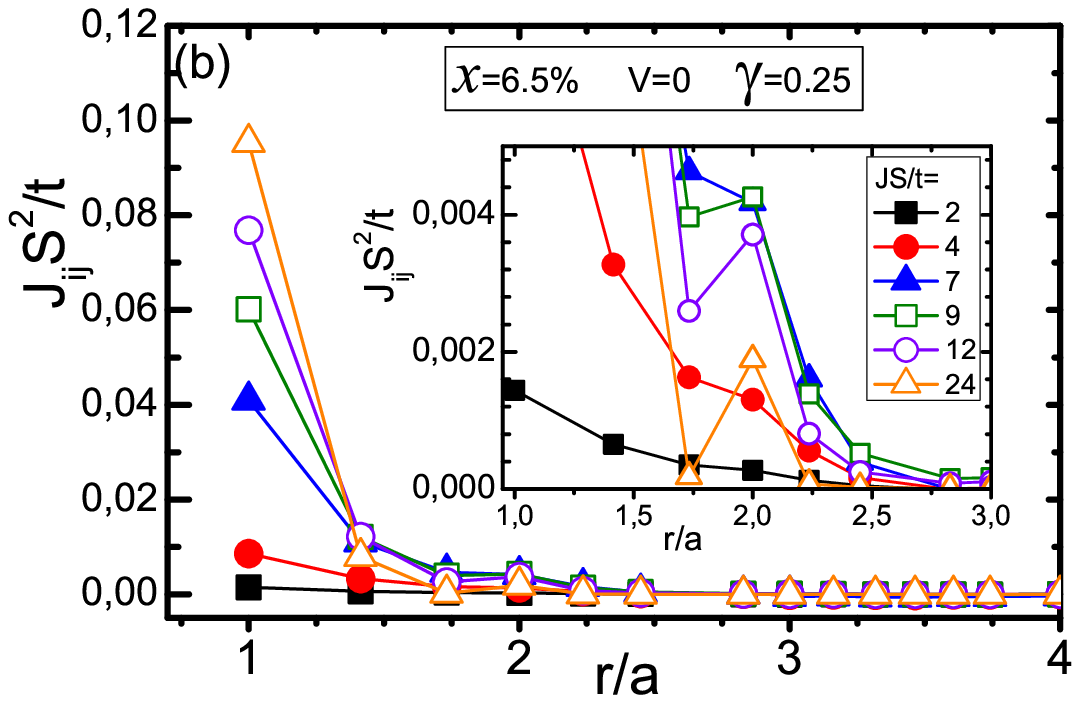}
\vspace{-0.70cm}
\caption{(Color online.) Magnetic couplings (in unit of t) as a function of the distance between magnetic impurities.
The density of impurity is $x=0.065$ and the hole concentration is respectively (a) $p_{h}=0.125 x$ (or $\gamma=0.125$) and (b) $p_{h}=0.25 x$ (or $\gamma=0.25$). The magnetic exchanges are calculated for different values of JS. In both figures, the inset focuses on the most ``relevant'' couplings.}
\label{Fig4}
\end{figure}
Fig.~\ref{Fig4} shows the couplings as a function of the impurity distance, for two different carrier concentrations and different values of JS. Note that, the calculations were performed on a sufficiently large cluster ($N=20^{3}$) to neglect the finite size effects. The discussion of the finite size effects will be done in a forthcoming paragraph. First, we observe in Fig.~\ref{Fig4} that the couplings are all ferromagnetic and of relatively short range. For $x=0.065$, the typical average distance between magnetic impurities is $\overline{d} \approx 2.4 a$. It corresponds to the distance between impurities on the ordered underlying super-lattice. For the smallest values of JS, the couplings are small at any distances, this should lead to very small Curie temperatures.
As JS is increased, the nearest neighbor coupling increases strongly until it saturates whilst the range of the couplings is reducing (clearly seen in the insets).
For very large JS (JS $\gg$ W), the couplings are essentially reduced to the nearest neighbor coupling only, as expected in the double exchange limit. The limiting value is expected to be independent of the carrier density and corresponds roughly to the kinetic energy of a single hole hopping between two sites. Note that there are also some additional contributions coming from one hole localized on three sites, 2 holes on three sites and so on. One can already say that, in the large JS limit, the Curie temperature is expected to vanish since the average distance between impurities is larger than the range of the couplings. In other words, the system is below the percolation threshold. For $JS \approx 2W$ (see DOS in fig.1), the couplings at distances near $\overline{d}$ are tiny, thus one expects very small Curie temperatures.

\begin{figure}[htbp]
\includegraphics[width=9.0cm,angle=-0]{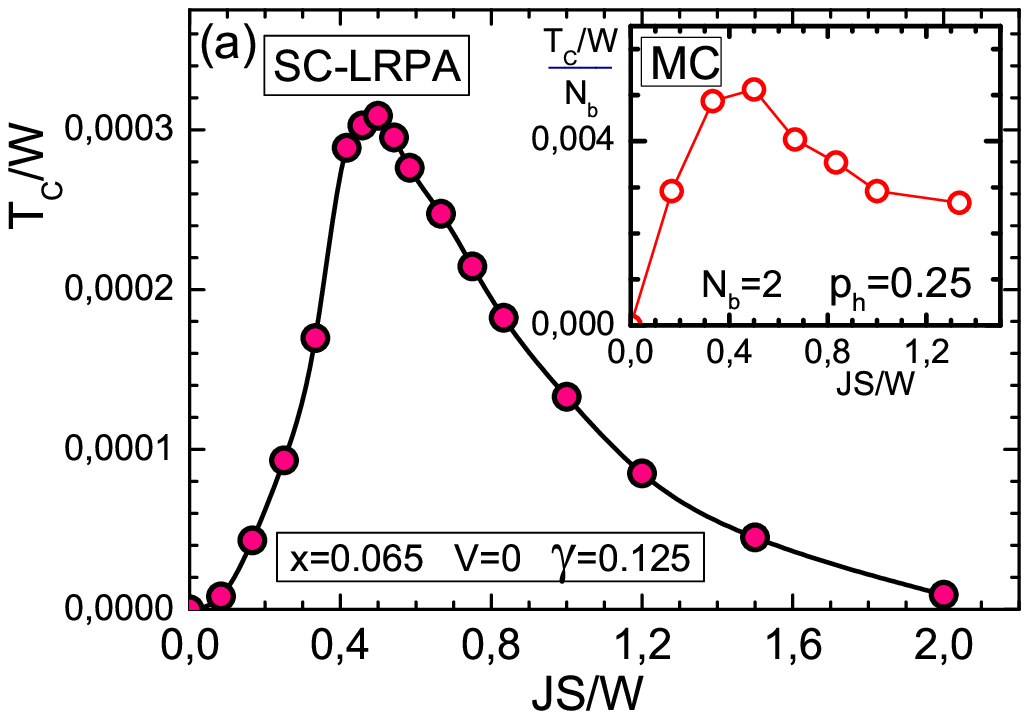}
\includegraphics[width=9.0cm,angle=-0]{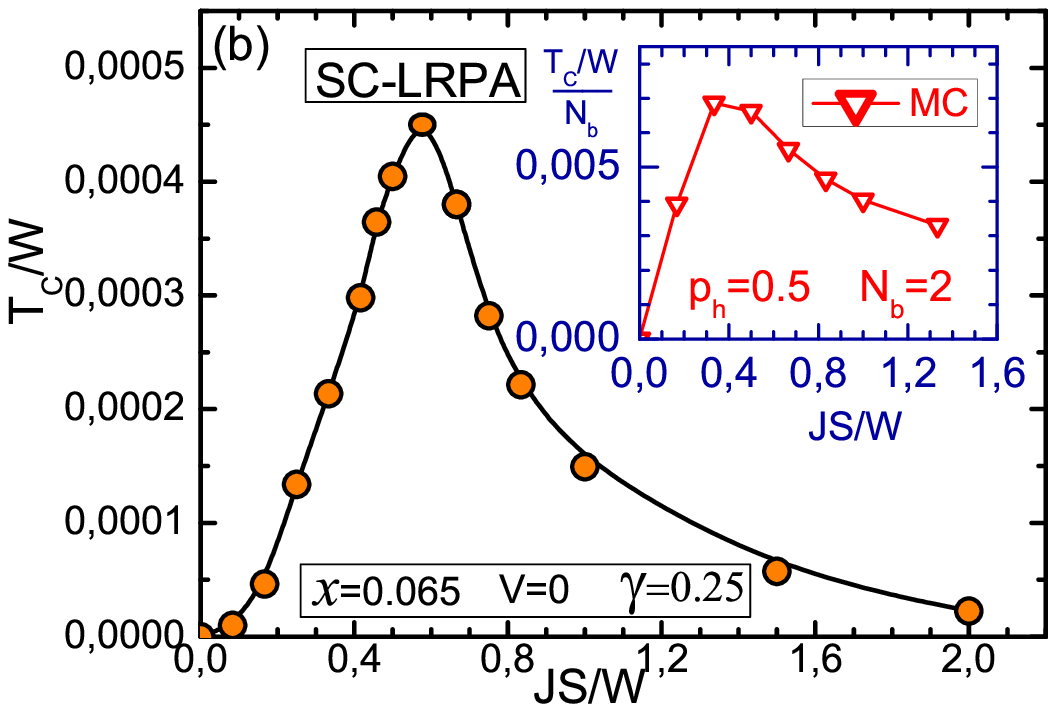}
\vspace{-0.0cm}
\caption{(Color online.)
Curie temperature within SC-LRPA (in units of the bandwidth W$=$12t) as a function of JS/W.
The magnetic impurity concentration is fixed $x=0.065$. The hole concentration per impurity is (a) $\gamma=0.125$ and (b) $\gamma=0.25$. The insets represent the Monte Carlo (MC) calculations from ref.\cite{dagottoetal} done for a
two band model ($N_{b}=2$) in which (a) $\gamma=0.5$  and (b) $\gamma=0.25$ }
\label{Fig5}
\end{figure}

\begin{figure}[htbp]
\includegraphics[width=9.0cm,angle=-0]{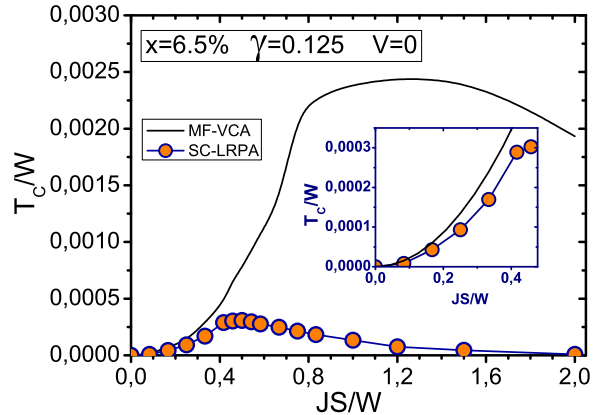}
\caption{(Color online.)
MF-VCA and full SC-LRPA calculations of the Curie temperature as a function of JS. The magnetic impurity concentration is $x=0.065$ and density of hole per carrier is $\gamma=p_{h}/x=0.125$.
}
\label{Fig6}
\end{figure}

In Fig.~\ref{Fig5}, we have plotted the Curie temperature as a function of JS. The carrier density is the same as used in Fig.~\ref{Fig4}. We again recall that the Monte Carlo treatment of this Hamiltonian would lead to similar results as it was already shown in several publications \cite{bergqvist,Bergqvist}. For both values of $\gamma$, we observe that T$_{C}$ increases until it reaches a maximum, then it decreases until it vanishes beyond a critical value which depends on the magnetic impurity concentration. Note that the maximum is located at $JS \approx 0.5$ for $\gamma=0.125$ and $JS \approx 0.6$ for $\gamma=0.25$.
The highest T$_{C}$ are respectively $T^{max}_{C}(\gamma=0.125) \approx 3. 10^{-4} ~W$ and $T^{max}_{C}(\gamma=0.25) \approx 5. 10^{-4} ~W$. For small values of J (perturbative limit) one notices that within SC-LRPA  $T_{C} \propto J^{2}$, as expected from Eq.(\ref{couplings}).
In Fig.~\ref{Fig6}, we evaluate the role of the disorder and the importance of the thermal fluctuations. For that purpose, we compare T$_{C}$ calculated within SC-LRPA to the Mean Field Virtual Crystal Approximation (MF-VCA) value. We remind that the MF-VCA value is T$^{MF-VCA}_{C}$$=$ $\frac{2}{3}x$  $\sum_{i \ne 0}$ J$_{0i}$. Within this approximation both thermal and transverse fluctuations and disorder are treated at the lowest order. In fact the disorder (dilution)is neglected and just appears as a trivial prefactor ($x$) in the expression of T$_{C}$. In the limit of the very small J (inset of Fig.~\ref{Fig6}), we observe a good agreement between SC-LRPA and MF-VCA. Within the perturbative limit, and for relatively weak densities of carriers only, one could indeed expect such a behavior (similar results are observed for $\gamma=0.25$ of Fig.~\ref{Fig5}a). As we increase JS, we observe important quantitative and qualitative differences, as T$^{MF-VCA}_{C}$ $\gg$ T$^{SC-LRPA}_{C}$.

\begin{figure}[tbp]
\vspace{-0.0cm}
\includegraphics[width=9.0cm,angle=-0]{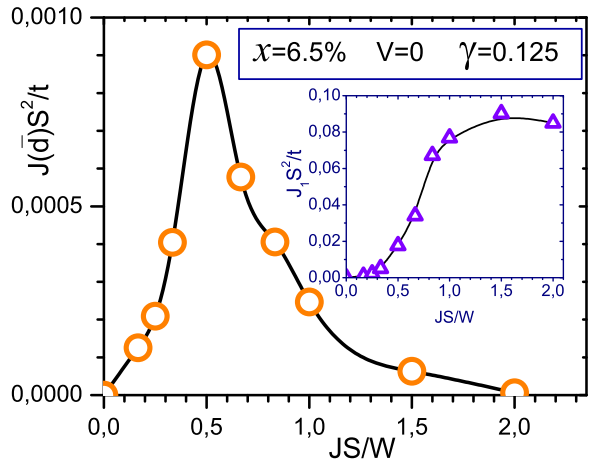}
\caption{(Color online.) $J(\overline{d})S^{2}$ in unit of t as a function of the local coupling JS/W. The density of holes is fixed and set to $p_{h}=0.125 x$. In the inset we have plotted the nearest neighbor coupling $J_{1}S^{2}$ as a function of JS/W.
}
\label{Fig7}
\end{figure}

Let us now show that T$_{C}$ is in fact mainly controlled by the typical coupling namely $J(\overline{d})$, where
$\overline{d}$ was defined previously. For $x=0.065$ we remind that the average distance between impurities is $d=\overline{d} \approx 2.4 a$. In Fig.~\ref{Fig7} we have plotted the variation of this coupling as a function of $JS/W$. We clearly see that the shape of the calculated curve is very similar to that obtained in Fig.~\ref{Fig5}a. We find that the ratio $R=\frac{T_{C}}{J(\overline{d})S^{2}}$ is almost independent of JS and $R \approx 4$. On the other hand, the non monotonic behaviour of $J(\overline{d})$ is qualitatively different from the monotonic increase of $J_{1}$ as seen in the inset. The limiting value of $J_{1}$ (for $JS \rightarrow \infty$) is expected to be close to that of the clean system as plotted in Fig.1. For large $JS$, $J_{1}$ dominates and leads to the unrealistic MF-VCA critical temperatures (see Fig.~\ref{Fig6}).

\section{Comparison with available full Monte Carlo calculations}

In this section, we propose to compare our results with recent full Monte Carlo simulations results available in the literature.
We will see that the quoted full Monte Carlo results are questionable since suffering from severe numerical shortcomings.
To our knowledge, the only available full Monte Carlo study of the diluted Hamiltonian given by eq. (\ref{Hamiltonian}) was performed by Popescu et al. \cite{dagottoetal}, no other group has performed similar studies in the dilute regime. In their MC simulations, Popescu et al. consider the case of two identical independent bands. Thus, both $T_{C}$ and the carrier densities used in their simulations are simply divided by two in order to allow a direct comparison with our calculations (performed on a single band model).
The MC data of ref. \cite{dagottoetal} are shown both in the insets of Fig.~\ref{Fig5}a and b.
The MC critical temperatures were obtained for systems with $N= 5^{3}$ to $6^{3}$ sites and the average over disorder
was performed over 7 configurations of disorder only. Let us notice that a system of size $N=5^{3}$ (resp. $6^{3}$) contains only 8 (resp. 14) magnetic impurities; thus for a density of carriers per impurity $\gamma=0.25$, such small systems respectively contain only 2 and 3 holes in the whole cluster.
As seen in Fig.~\ref{Fig5}, several crucial differences appear immediately with respect to our results. First, we notice that for both carrier densities ($\gamma=$ 0.125 and 0.25) the Monte Carlo simulations predict critical T$_{C}$ much higher than those obtained within the two steps approach. Indeed, if one refers for example to Fig.~\ref{Fig5}a, one observes that the MC values are 15 to 60 times larger! In particular, the highest $T_{C}$ are typically at least 20 times larger than our results. For small values of JS, one notices that SC-LRPA and the Monte Carlo simulations do not predict the same behavior (quadratic in J in our case).

For large JS, whilst our calculated T$_{C}$ is strongly suppressed and vanishes for values of $JS \ge J_{c}S$ where $J_{c}S \approx 2~W$, the MC value decreases much slowly and seems to saturate at large JS (see Fig.~\ref{Fig5}a). For example, for $\gamma$$=$0.125 and JS$=$1.2~W  (or p$_{h}$$=$0.25 for the 2 bands MC calculations), $T^{MC}_{C}/N_{b}= 3\cdot10^{-3}~W$ whilst we find $T^{LRPA}_{C}= 7.5 10^{-5} ~W$, thus the ratio $T^{MC}_{C}/T^{LRPA}_{C}=40$, which is huge! Additionally, there is an other unexpected and surprising result. The MF-VCA Curie temperature which should in principle be an upper bound is in fact {\bf smaller} than the Monte Carlo value. This is in conflict with the opposite and natural expectation. We recall that the Monte Carlo method treats both disorder effects and thermal fluctuations exactly. For example, from Fig.~\ref{Fig6} and for JS$=$0.4~W, we find T$^{MF-VCA}_{C}$$=$ 4 $\cdot$ 10$^{-4} ~W$ and T$^{MC}_{C}$$=$ 50 10$^{-4}$ $~W$. Additionally, it was already shown several times in the literature, especially for the study of the diluted magnetic semiconductors, that T$^{MF-VCA}_{C}$ largely overestimates the real Curie temperature, this clearly put a question mark on the validity of the full MC results.

On the basis of what was previously discussed, we now address the reasons which may explain the huge differences with TSA and why the Mean Field VCA becomes a lower bound with respect to the full Monte Carlo results. To be more specific, it will be shown that the available Monte Carlo results suffer from several combined effects: (i) strong finite size effects, (ii) insufficient sampling (too few disorder configurations) and (iii) inaccurate and approximative procedure for the determination of the Curie temperature.

\begin{figure}[tbp]
\vspace{-0.0cm}
\includegraphics[width=9.0cm,angle=-0]{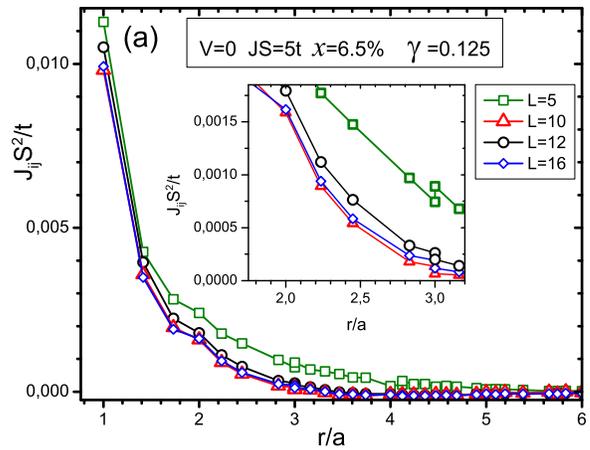}
\includegraphics[width=9.0cm,angle=-0]{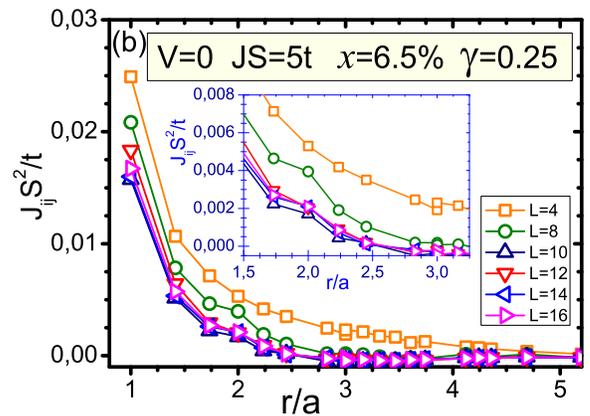}
\vspace{-0.70cm}
\caption{(Color online.) Magnetic couplings $J_{ij}(r)S^{2}$ in unit of t as a function of the the distance between impurities r/a.The magnetic impurity concentration is fixed $x=0.065$, JS$=$5~t and (a) $\gamma=0.125$ and (b) $\gamma=0.25$ . The calculations are done with clusters of size $N=L^{3}$ where L$=$ 4,8,10,12,14,16.}
\label{Fig8}
\end{figure}

\section{Origins of the dissensions between TSA and Monte Carlo simulations}

\begin{figure}[htbp]
\includegraphics[width=9.0cm,angle=-0]{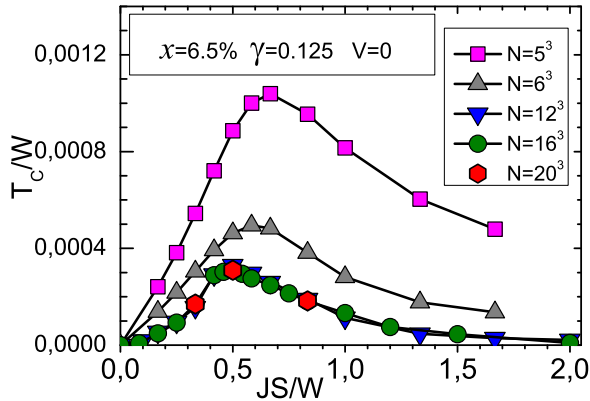}
\caption{(Color online.) Averaged Curie temperature (in units of the bandwidth W) as a function of the parameter JS/W.
The magnetic impurity concentration is fixed $x=0.065$. The hole concentration per impurity is $\gamma=0.125$.
 The calculations are done with clusters of size $N=L^{3}$ where L$=$ 5,6,12,16,20. A systematic average over few thousands of configurations have been done for the smaller systems and few hundreds for the largest.
}
\label{Fig9}
\end{figure}

In this section we will discuss both the importance of the statistical sampling and finite size effects.
In Fig.\ref{Fig8} a and b ($\gamma=0.25$ and 0.125) we have plotted the variation of $J_{ij}S^{2}$ as a function of the distance between impurities, for different system sizes ($N=L^{3}$ where L varies from 4 to 16). We observe that the couplings are larger for the smallest systems and that finite size effects are especially strong at distances near the typical and relevant one, $\overline{d}$. Thus, we already expect the Curie temperature to be strongly size dependent. In Fig.\ref{Fig9} the average Curie temperature as a function of $JS/W$ is shown for various sizes. The hole density (per carrier) is set to $\gamma=0.125$. Note that each averaged T$_{C}$ is obtained using the associated couplings, e.g calculated for the corresponding system. Note also that the average was properly performed, we have used a sufficiently large number of disorder configurations (few thousands for the smallest systems and few hundreds for the largest). We analyze now the importance of the statistical sampling. For a given value of JS, we observe beyond $N=14^{3}$ an insignificant variation of the averaged Curie temperature: beyond this size the thermodynamic limit is properly described.
Because within standard full Monte Carlo method the systems are usually restricted to relatively small sizes \cite{dagottoetal,Popescu07,Yildirim07,Alvarez02,Alvarez03} of the order of $N=4^{3}$, let us do the comparison between the calculations done with the smallest and largest systems. For instance, we consider $JS=0.7~W$. We see that the Curie temperature for $L=5$ is $T_{C}=10.5~10^{-4}~W$ whilst in the thermodynamic limit $T_{C}=2.0~10^{-4}~W$ (500 \% difference!).

We now show that the sampling over disorder is also crucial and should be done properly, especially in dilute systems. Note for instance that, Franceschetti et al. \cite{Franceschetti06}, in their ab initio based study, have shown important fluctuations of the Curie temperature from sample to sample.
\begin{figure}[htbp]
\includegraphics[width=8.50cm,angle=-0]{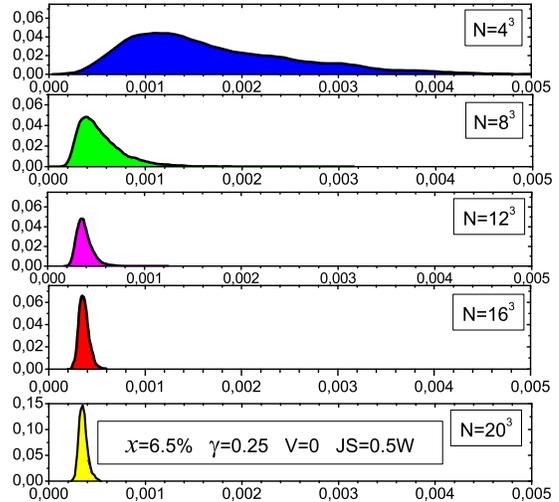}
\caption{(Color online.) Distribution of the calculated Curie temperatures for different systems sizes N$=$$L^3$ where L varies from 4 to 20. The parameters are $x=0.065$, $V=0$, $\gamma=0.25$ and $JS=0.5~W$
}
\label{Fig10}
\end{figure}
In Fig.\ref{Fig10} we have plotted the distribution of the Curie temperatures obtained using different system sizes.
To facilitate the discussion, we have plotted on Fig.\ref{Fig11} the averaged Curie temperatures, $\overline{T^{L}_{C}}=\int_{-\infty}^{+\infty} T_{C}P_{L}(T_{C})dT_{C}$, and the widths at mid height of the distributions ($\Delta T_{C} /W$) as a function of 1/N. For the smallest system ($N=4^{3}$) one observes a very important spreading of the critical temperatures distribution: they can vary from at least one order of magnitude! From Fig.\ref{Fig11}, for this case, $\Delta T_{C} /W$  is rather close to the corresponding average value. For these small systems, one understands easily that 10 configurations of disorder are definitely insufficient to provide a reliable average.
In other words, the calculated averaged Curie temperature over a small number of configurations (few tens) could easily lead to critical temperatures 10-20 times larger than that calculated properly (averaged over few thousands of configurations). This implies, for example, that the Curie temperatures shown in Fig.\ref{Fig9} could be easily 50 times larger than the calculated ones in the thermodynamic limit if the average would have been taken over only few configurations.

\begin{figure}[htbp]
\includegraphics[width=8.50cm,angle=-0]{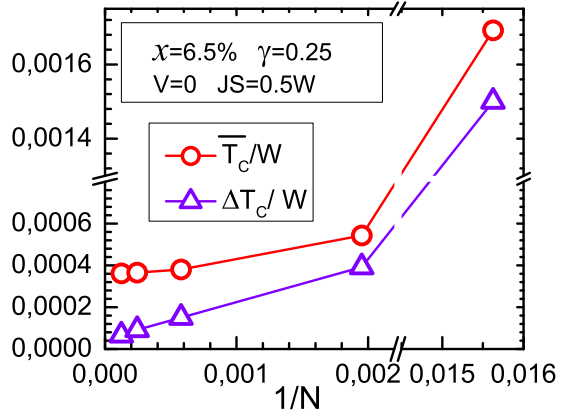}
\caption{(Color online.)
Width ($\Delta T_{C}$) and average ($\bar{T_{C}}$) of the Curie temperatures distribution calculated for various
simple cubic lattices of size N$=$L$^{3}$ as a function of 1/N. The parameters
are the same as those of Fig.\ref{Fig10}.}
\label{Fig11}
\end{figure}
 As the system size increases, one gradually observes a decrease in $\Delta T_{C}$, which then rapidly tends to zero (thermodynamic limit), whilst $\overline{T^{L}_{C}}$ tends towards a constant value (see Fig.\ref{Fig11}). One can already consider that the thermodynamic limit value of $T_{C}$ is reached for $L\geq16$. In the light of the results presented in this manuscript, a criterion to get reasonable results would consist in considering system sizes where L is at least five to six times larger than the typical distance between impurities. This criterion is only rough, in the case of long range magnetic couplings, such as RKKY exchanges, the finite size effects can be even more drastic. In addition, only ten configurations of disorder for small systems are definitely insufficient to get the average value of $T_{C}$ in a reliable manner. It is obvious that these numerical requirements are very difficult to fulfill within standard Monte Carlo calculations but they are definitely essential.
In addition, in the presence of inhomogeneities, temperatures are also expected to be even more sensitive to both finite size effects and statistical sampling \cite{RBouzerar08inhomoinprep}. Moreover, the way of extracting the Curie temperature is essential. Within SC-LRPA, the problem does not arise. The critical temperatures are directly given by a semi-analytical equation solved in a self-consistent way, no extrapolation from the magnetization curve is used. Concerning the Monte Carlo simulations of reference \cite{dagottoetal}, the method used to extract $T_{C}$ from the magnetization curve is not very accurate (see figure 4.a and 4.b of ref. \cite{dagottoetal}). Potential errors related to this method are added to the uncertainties coming from both finite size effects and from the insufficiency of statistical sampling. A more accurate way to obtain the Curie temperatures in MC simulations is based on the method of Binder cumulants \cite{Binder02}, but this method represents an additional cost in terms of computing time since it needs a finite size effect analysis. Unfortunately, in spite of the small system sizes and the small number of configurations of disorder considered in ref.\cite{dagottoetal}, the Monte Carlo simulations already needs a huge resource in term of CPU time and memory.

In a recent paper based on Monte Carlo simulations, Yildirim et al. have studied the diluted Kondo model applied to the case of Ga$_{1-x}$Mn$_x$As by including the realistic band structure of the host material \cite{Yildirim07} (see also ref. \cite{GB-RB-comment08} for a comment).
They have calculated the Curie temperature as a function of the local coupling J for systems containing  $x= 8.5\%$ of magnetic impurities and $p_{h} = 0.75$ hole per impurity. We underline here that their numerical Monte Carlo approach suffers from the same numerical shortcomings as those previously discussed. The system considered in this letter contain typically 20 localized spins only and the average are done using just five configurations of disorder.
Additionally, the value of J considered in ref.\cite{Yildirim07}, corresponds to the perturbative regime. Thus, on the basis of the present study, we argue that within this limit, the couplings should exhibit RKKY oscillations which should lead in the thermodynamic limit to small Curie temperatures or eventually to no ferromagnetic phase \cite{RichardPRB}. Unfortunately, and as previously noticed, the smallness of the cluster considered in the MC calculations hide the effects of the asymptotic RKKY tail, and thus leading to finite and large Curie temperatures. We argue that by improving the statistics and increasing the size of the systems the Curie temperature should vanish when the frustration will become effective.
Furthermore, we argue that the large MC critical temperatures found in the large JS regime (see Fig.1 of ref. \cite{Yildirim07}) are in fact a numerical artefact due again to the insufficient statistical sampling and finite size effects. Indeed, in this regime double exchange mechanism dominates and thus very small or vanishing Curie temperatures are expected, see the corresponding densities of states in Fig.4 of \cite{Yildirim07} where the IB is separated from the valence band.

\section{Conclusion}

To conclude, in this work we have shown that the study of the diluted magnetic systems requires a rigorous numerical treatment. We have shed light on the origins of the disagreements between MC and TSA for the study of diluted magnetic systems. In contrast to the non dilute systems, the finite size effects as well as the importance of the statistical sampling appear to be crucial. We have shown that available Monte Carlo simulations for the diluted model Hamiltonian (\ref{Hamiltonian}) suffer from severe numerical insufficiencies. Although Monte Carlo simulations are in principle exact, but because the calculations are unfortunately restricted to relatively small systems and weak statistical sampling, the obtained Curie temperatures are often largely overestimated. In addition, limiting regimes (strong and weak couplings) are not properly described. 
It would be of great interest to check both the importance of statistical sampling and the finite size effects within new large scale Monte Carlo studies. Among MC simulations which allow to reach relatively large system sizes, one can quote the Hybrid Monte Carlo method (HMC) \cite{Alonso01HMC}, the Polynomial Expansion Monte Carlo (PEMC) \cite{Motome99_00_01_03PEMC} or its fastest counterpart, namely the Truncated Polynomial Expansion Monte Carlo (TPEMC) method \cite{Motome03TPEM,Furukawa04TPEM,Sen06TPEM}. In Particular TPEMC handles higher systems up to $20^{3}$ sites within a reasonable CPU time (see table II of ref. \cite{Furukawa04TPEM}).

\acknowledgments
First we would like thank E. Kats for the hospitality at ILL and for providing us
 with the access to the ILL theory computer facilities. We also thank Mark Johnson for
providing us the access to the AMD and Fujitsu Siemens clusters, and Mark Ash for
the access to MECS cluster. We also thank B. Barbara, O. C\'epas, C. Delerue, P. Qu\'emerais, D. Feinberg, E. Kats,
P. Bruno, J. Kudrnovsky and L.Bergqvist for interesting and stimulating discussions and remarks.

\end{document}